\documentclass[fleqn,twoside]{article}
\usepackage{espcrc2}
\usepackage{epsfig}
\usepackage{amssymb}
\usepackage{graphicx}
\usepackage[figuresright]{rotating}

\newcommand{\AmS}{{\protect\the\textfont2
  A\kern-.1667em\lower.5ex\hbox{M}\kern-.125emS}}

\newcommand{\msbar}{\overline{\mbox{\scriptsize MS}}}

\newcommand{\chil}{{\mbox{\scriptsize L}}}
\newcommand{\chir}{{\mbox{\scriptsize R}}}

\newcommand{\Tr}{\mbox{Tr}\;}
\newcommand{\<}{\langle}
\renewcommand{\>}{\rangle}

\newcommand{\simge}{\ \lower-
1.2pt\vbox{\hbox{\rlap{$>$}\lower5pt
\vbox{\hbox{$\sim$}}}}\ }

\newcommand{\bc}{\begin{center}}
\newcommand{\ec}{\end{center}}
\newcommand{\be}{\begin{equation}}
\newcommand{\ee}{\end{equation}}
\newcommand{\ba}{\begin{eqnarray}}
\newcommand{\ea}{\end{eqnarray}}
\newcommand{\brr}{\begin{array}}
\newcommand{\err}{\end{array}}

\title{Exact Chiral Symmetry on the Lattice: QCD 
       Applications\thanks{This work was supported in part by the 
                 European Community's Human Potential Programme under 
                 contract HPRN-CT-2000-00145, Hadrons/Lattice QCD.}}

\author{Leonardo Giusti
        \address[CERN]{CERN, Theory Division, 
                       CH-1211 Geneva 23, Switzerland}$^,$\address[MARSEILLE]
{Centre de Physique Th\'eorique, CNRS Luminy, Case 907, 
 F-13288 Marseille, France}}
\begin{document}

\begin{abstract}
\noindent I review recent progress and results in lattice QCD obtained using 
fermions with exact chiral symmetry. 
\vspace{1pc}
\end{abstract}

\maketitle

\section{Introduction}
In the last few years it has been understood that exact chiral symmetry 
can be realized at finite lattice spacing.
A breakthrough started with the observation that massive Dirac 
fermions in $4+1$ dimensions reduce to chiral fermions in $4$ dimensions
under certain conditions \cite{Rubakov:bb,Callan:sa,Kaplan:1992bt}.
The resulting effective operator $D$ of light boundary fields \cite{NaraNeub,Neuberger}
has the correct continuum limit, no doublers, and is local~\cite{Hernandez:1998et}. 
Most remarkably, it satisfies the Ginsparg--Wilson (GW) relation \cite{Ginsparg:1982bj}
\be\label{eq:GW}
\gamma_5 D + D \gamma_5 = \bar a D \gamma_5 D \; ,  
\ee
which guarantees an exact chiral symmetry \cite{Luscher:1998pq}
\be
\delta q = \gamma_5(1-\bar a D) q\; , \qquad
\delta \bar q = \bar q \gamma_5  
\ee
of the fermion action at non-zero lattice spacing.

Within the perfect-action approach \cite{PerfectA}, a fermion operator 
that satisfies the GW relation can be defined 
\cite{Hasenfratz:1997ft}, but no explicit construction has been found 
so far (see \cite{gattringer} for a recent review).

In QCD a chiral symmetric regularization entails many theoretical 
advantages: in the infrared it allows one to simulate massless quarks
and it provides a natural definition of the topological charge; 
in the ultraviolet it simplifies 
the subtraction of divergences in composite operators.

Last year an important step forward was the demonstration that 
Neuberger's fermions can be used for large-scale QCD computations, 
at least in the quenched approximation. As a result, very light quarks can be handled 
and legendary problems such as the $\Delta I=1/2$ rule in 
$K\rightarrow\pi\pi$ decays are greatly simplified and can be attacked.

In this talk some of the phenomenological applications
of Ginsparg--Wilson fermions are reviewed. Most
of the numerical studies reported are exploratory.
Please refer to the parallel session contributions
for interesting theoretical developments which are not  
covered in the following.

\section{Domain-wall-overlap fermions}
The five-dimensional domain-wall Dirac operator 
can be defined as \cite{Kaplan:1992bt,Shamir:1993zy}
\be
{\cal D} = \frac{1}{2}\Big[\gamma_5(\partial^*_s + \partial_s) - 
a_s \partial^*_s\partial_s\Big] + X \; ,
\ee
where $s$ labels the sites and $a_s$ the lattice 
spacing in the fifth dimension. The operators $\partial^*_s$ and 
$\partial_s$ are the forward and backward derivatives, 
\be
X = D_W - \frac{1}{\bar a} \qquad , \qquad \bar a =\frac{a}{\rho}\; ,
\ee
with $0<\rho<2$.  The massless four-dimensional Wilson  operator
is 
\be
D_W = \frac{1}{2} \Big[\gamma_\mu (\nabla_\mu + \nabla^*_\mu)
- a \nabla^*_\mu\nabla_\mu\Big]\; , 
\ee
where $\nabla_\mu$ and $\nabla^*_\mu$ are the gauge-covariant forward 
and backward derivatives 
\ba
\nabla_\mu q(x) & = & \frac{1}{a}\Big[U_\mu(x) q(x + a \hat \mu) -
q(x)\Big]\label{eq:derivative}\\
\nabla_\mu^* q(x) & = & 
\frac{1}{a}\Big[ q(x) - U^{\dagger}_\mu(x-a \hat \mu) 
q(x-a\hat\mu) \Big]\; \nonumber 
\ea
and $a$ is the lattice spacing in the physical four dimensions. 
The operator is supplemented with the boundary conditions
\be
P_+ q(0,x) =  P_- q(a_s N_s + a_s,x) =  0
\ee
where $P_{\pm} = \frac{1}{2}(1\pm \gamma_5)$ and 
$N_s$ is the extension in the fifth dimension. 
From a four-dimensional point of view, the system corresponds to QCD with many flavours 
mixed in a peculiar way. 

By integrating out the heavy flavours, a four-dimensional effective action of 
light boundary fields is left \cite{NaraNeub,Neuberger}
\be
\bar{a} D_{N_s} = 1 + \gamma_5 \frac{(1+\tilde H)^{N_s} - (1-\tilde H)^{N_s}}
                             {(1+\tilde H)^{N_s} + (1-\tilde H)^{N_s}}\; ,
\ee
where 
\be
\tilde X \equiv \frac{a_s X}{2+a_s X} \qquad , \qquad  \tilde H \equiv \gamma_5 \tilde X \; .
\ee
For $N_s\rightarrow\infty$, a massless effective action
\be\label{eq:walls}
\bar{a} D_{DW} = \left( 1 + \tilde X\frac{1}{\sqrt{\tilde X^\dagger \tilde X}}\right)\nonumber 
\ee
is derived and the Neuberger  operator \cite{Neuberger} 
\be\label{eq:neub}
\bar{a} D_N = \left( 1 + X\frac{1}{\sqrt{X^\dagger X}}\right)\nonumber
\ee
is obtained if the limit $a_s\rightarrow 0 $ is also taken. 
Most remarkably, 
the effective massless Dirac 
operators in Eqs.~(\ref{eq:walls}) and (\ref{eq:neub}) 
satisfy the GW relation \cite{Neuberger:1998wv}.

\section{Exact chiral symmetry at finite $a$}
For a given operator $D$ that satisfies the GW relation,
the QCD fermion action with $N_f$ flavours can be written as 
\ba\label{eq:action}
\frac{S_F}{a^4} = \sum_x \bar{\psi}(x) \left[(D + P_- {\cal M}^\dagger \hat P_- + P_+ {\cal M} \hat P_+ 
)\psi \right](x)\nonumber
\ea
where 
\be
\hat P_{\pm} = \frac{1}{2}(1\pm \hat \gamma_5)\; , \qquad \hat \gamma_5=\gamma_5(1-\bar a D)\; ,
\ee
${\cal M}=\mbox{diag}(m_1,\dots , m_{N_f})$,
$\bar \psi=(\bar q^{_1},\cdots , \bar q^{_{N_f}})$ and $\psi$ is defined analogously.
It is invariant under the U$(N_f)_{\chil}\times$U$(N_f)_{_\chir}$
global transformations
\ba\label{eq:chitransf}
\psi_\chil \rightarrow V_\chil \psi_\chil\;  & \qquad & 
\bar \psi_\chil \rightarrow \bar \psi_\chil V_\chil^{\dagger}\nonumber\\
\psi_\chir \rightarrow V_\chir \psi_\chir\;  & \qquad & 
\bar \psi_\chir \rightarrow \bar \psi_\chir V_\chir^{\dagger}\; ,
\ea
where $V_{\rm L,R}\in$U$(N_f)_{\chil,\chir}$ and 
\be
\psi_{\chir,\chil} = \hat P_{\pm} \psi \qquad \bar\psi_{\chir,\chil} = \bar\psi P_{\mp}\; ,
\ee
if also ${\cal M}\rightarrow V_\chil {\cal M} V_\chir^\dagger$.
No additive quark mass renormalization is required.
The action is $O(a)$-improved, since no chiral invariant 
operators of dimension $d=5$ can be constructed. 

The global chiral anomaly is recovered \`a la Fujikawa 
\cite{Fujikawa:1979ay,Luscher:1998pq}. The fermion integration 
measure is not invariant under $U(1)_A$ transformations, and 
the topological charge density from the corresponding Jacobian
\be
a^4 Q(x) = \frac{\bar a}{2 a} \Tr \Big[\gamma_5 D(x,x)\Big]\; 
\ee 
satisfies \cite{Neuberger,Hasenfratz:1998ri}
\be
n_- - n_+={\rm{index}}(D)=a^4 \sum_x \, Q(x) \; , 
\ee
with $n_\pm$ the number of right and left zero modes of the 
fermion operator.

Bilinear fermion operators with proper chiral transformations 
\be\label{eq:bilinears}
{\cal O}_\Gamma^{\alpha\beta}(x) =\bar q^\alpha(x) \Gamma \tilde q^\beta(x) 
\, , \;\; \tilde q^\beta = \Bigl(1-\frac{\bar a}{2}D\Bigr) q^\beta
\ee
are $O(a)$-improved, but they do not transform in a 
simple way under CP \cite{kpipi_club,Fujikawa:2002vj}.
However in correlation functions of local operators at non-zero 
physical distance, it holds 
\be
{\cal O}_\Gamma^{\alpha\beta}(x) = 
\frac{1}{(1-\frac{\bar a}{2} m_\beta )} \, \bar q^\alpha(x) 
\Gamma q^\beta(x) 
\ee
and a simple CP transformation [$\tilde x = (x_0,-\vec{x})$]
\be
{\cal O}_\Gamma^{\alpha\beta}(x) \stackrel{\mbox{CP}}{\longrightarrow}  
\frac{1-\frac{\bar a}{2} m_\alpha}{1-\frac{\bar a}{2} 
m_\beta} \, {\cal O}_\Gamma^{\beta\alpha}(\tilde x) 
\ee
is recovered \cite{kpipi_club}. The generalization 
to four-fermion operators is straightforward.

Non-singlet local rotations lead to exact vector and axial Ward identities (WI),
and the very same definition of the bare quark mass appears in 
the axial and vector WIs and in the quark propagator. 

The conserved currents ${\cal V}^a_\mu$ and ${\cal A}^a_\mu$ can be constructed 
by extending the gauge group $SU(N_c)\rightarrow SU(N_c)\times U(1)$
\cite{Ginsparg:1982bj,Kikukawa:1998py}. 
By performing a local $U(1)$ flavor rotation
\be
U_\mu(x) \rightarrow U_\mu^{(\alpha)}(x) = e^{i\alpha_\mu(x)} U_\mu(x) \; ,
\ee
the kernel 
\be
K_\mu = -i \frac{\delta D(U_\mu^{(\alpha)})}{\delta \alpha_\mu(x)}\Bigr|_{\alpha=0}
\ee
can be used to define the following conserved currents \cite{Kikukawa:1998py}
\ba
{\cal V}^{a}_\mu & = & \bar q 
\Bigl(P_- K_\mu \hat P_+ + P_+ K_\mu \hat P_- \Bigr) T^a q\\
{\cal A}^{a}_\mu & = & \bar q 
\Bigl(P_- K_\mu \hat P_+ - P_+ K_\mu \hat P_-\Bigr) T^a q
\ea
where $T^a$ is a generator of the non-singlet transformations.
It is interesting to note that (conserved) current--current correlators and 
their generalizations require the propagator from any point 
to any point \cite{kpipi_club}.

In the chiral limit, the local anomalous flavour-singlet WIs read
\be
\<\partial^*_\mu {\cal A}^0_\mu (x) \hat{\cal O}\>=2N_f \< Q(x) \hat{\cal O}\>+
\<\delta^x_A \hat{\cal O}\> \, , 
\label{AWTIL} 
\ee
where $\delta^x_A \hat{\cal O}$ is the local variation of 
any finite (multi)local operator $\hat{\cal O}$, and ${\cal A}^0_\mu (x)$ is 
the singlet axial current. By assuming the absence of a 
$U_A(1)$ massless Goldstone boson, the corresponding integrated WIs read
\be
0=2 N_f a^4 \sum_x  \<Q(x) \hat{\cal O}\> + \<\delta_A \hat{\cal O}\>\, . 
\label{AWTI} 
\ee
Since the second term in the r.h.s. of Eq.~(\ref{AWTI}) is  
finite, it follows that $a^4 \sum_x  Q(x)$ is also finite, 
as it has finite insertions with any string of renormalized  
fundamental fields. Therefore $Q(x)$ can only mix with  
operators of dimension $\leq 4$ and vanishing integral, hence only with  
$\partial^*_\mu {\cal A}^0_\mu(x)$. 
{\it No power-divergent subtractions} with lower dimensional operators 
(such as the pseudo-scalar quark density) have to be performed \cite{Giusti:2001xh}.
This is a very distinctive feature of GW fermions.
Calling $Z$ the mixing coefficient,  
one can define finite operators $\hat Q$ and $\hat {\cal A}^0_\mu$ by writing 
\be 
\hat{Q}(x)=Q(x) - \frac{Z}{2N_f}\partial^*_\mu {\cal A}_\mu^0 (x) 
\label{FINQ1}
\ee
\be
\hat{\cal A}_\mu^0 (x)=(1-Z) {\cal A}_\mu^0 (x)\, , 
\label{FINA} 
\ee 
and the renormalized singlet axial WIs are 
\be 
\<\partial^*\hat {\cal A}^0_\mu (x) \hat{\cal O}\>= 
2N_f\<\hat Q(x) \hat{\cal O}\>+\<\delta^x_A \hat{\cal O}\> \, . 
\label{WTIREN} 
\ee 

Renormalization constants of several composite operators
have been studied at one loop in perturbation theory for  
domain-wall fermions \cite{Aoki:1997xg,Aoki:1998vv,Aoki:1998hi,Aoki:1999ky,Aoki:2000ee,Aoki:2002iq} and 
overlap fermions 
\cite{Ishibashi:1999ik,Alexandrou:1999wr,Alexandrou:2000kj,Capitani:2000aq,Capitani:2000da,Capitani:2000bm,DeGrand:2002va}. 
Non-perturbative determinations have been
obtained for the non-singlet local vector and axial 
currents \cite{Blum:2001sr,Giusti:2001pk,Dong:2001fm}, for 
the scalar and pseudoscalar densities \cite{Blum:2001sr,Giusti:2001pk,Hernandez:2001yn}, and for 
some four-fermion operators \cite{Blum:2001xb,Marseille}.

\section{Meson spectroscopy}
\begin{figure*}[htb]
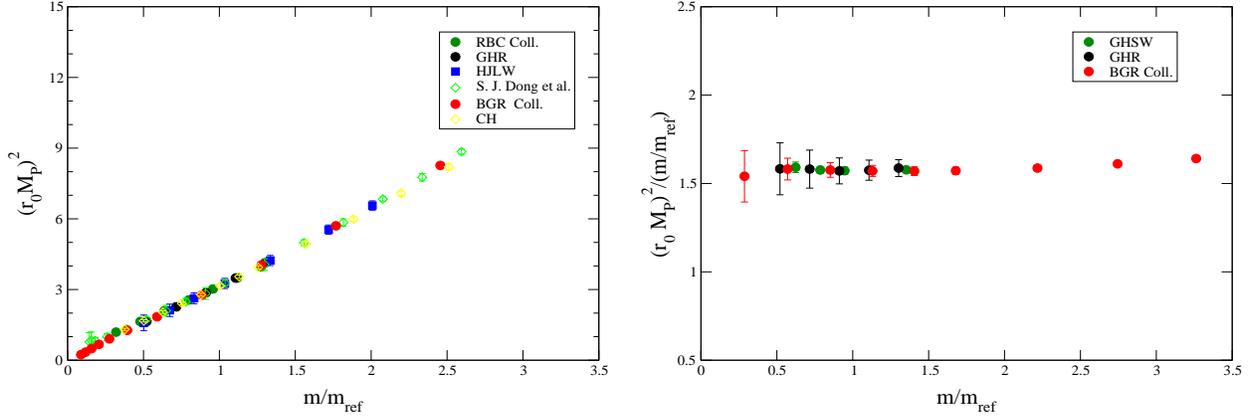

%\vspace{-1.0cm}
\catcode`?=\active \def?{\kern\digitwidth}
%\begin{center}
\hspace{-0.6cm}\begin{tabular}{cc}
\mbox{\epsfig{file=mp2_figure.eps,width=5.5cm,height=8.0cm,angle=270}} &
\mbox{\epsfig{file=mp2_mq_figure.eps,width=5.5cm,height=8.0cm,angle=270}} \\
\end{tabular}
\vspace{-0.8cm}
\caption{On the left, $(r_0 M_P)^2$ vs the quark mass normalized at the reference point 
$M^2_P=2M^2_K$: green circles \cite{Blum:2000kn},  black circles \cite{Giusti:2001pk}, 
blue squares \cite{Hernandez:2001yn},
green diamonds \cite{Dong:2001fm}, red circles \cite{Gattringer:2002sb}, yellow diamonds \cite{Chiu:2002xm}.
On the right, $(r_0 M_P)^2/(m/m_{\rm ref})$ vs $(m/m_{\rm ref})$: green circles \cite{Garden:1999fg}, black circles 
\cite{Giusti:2001pk},
red circles \cite{Gattringer:2002sb}.
\label{fig:moltobella}}
%\end{center}
\end{figure*}

In the past year several collaborations have computed the 
meson spectrum and light quark masses using GW fermions, 
either with the overlap formulation 
\cite{Giusti:2001pk,Dong:2001fm,Hernandez:2001yn,Chiu:2002xm} or with the 
perfect actions \cite{Hasenfratz:2002rp,Gattringer:2002sb}. 
Simulated lattices have linear extensions 
$L=1$--$3$~fm and lattice spacings $a=0.08$--$0.2$~fm.
Results for the pion mass squared in units of $1/r^2_0$ ($r_0=0.5$ fm)
as a function of the quark mass normalized  at the reference point 
$M^2_P=2M^2_K$ ($M_K=495$ MeV), are shown in the first plot of Fig.~\ref{fig:moltobella}.
Data in the range $500 \lesssim M_P \lesssim 800$ MeV show a linear behaviour
with a vanishing intercept within the statistical errors. 
The linearity manifests itself as a wide plateau in the 
second plot of Fig.~\ref{fig:moltobella}
and it is in very good agreement with what was previously observed with Wilson-type 
fermions in the same range of masses 
(see for example \cite{Allton:1996yv,Garden:1999fg,Aoki:2002fd}).
For degenerate quarks, quenched chiral perturbation theory at the next-to-leading order 
(NLO) predicts \cite{Colangelo:1997ch,Heitger:2000ay}
\ba\label{eq:pappante}
& & \hspace{-0.4cm} \displaystyle{\frac{M_P^2}{2 m}}  =  \displaystyle{\frac{\Sigma}{F^2}}
\Bigl[1-\delta\Bigl(1 + \log{\Bigl(\frac{M^2}{\mu_\chi^2}\Bigr)}\Bigr) + \\
& & \hspace{-0.8cm} \displaystyle{\frac{\alpha M^2}{3 (4\pi F)^2}}
\Bigl(1+2\log{\Bigl(\frac{M^2}{\mu_\chi^2}\Bigr)}\Bigr)
+  \Bigl(2 \alpha_8 -\alpha_5 \Bigr) 
\frac{M^2}{(4\pi F)^2}  \Bigr] \nonumber
\ea
where $\Sigma$, $F$, $m_0$, $\alpha$ are the leading-order (LO)
couplings of the quenched QCD chiral Lagrangian \cite{Bernard:1992mk,Sharpe:1992ft}, 
$\alpha_5,\alpha_8$ are some of the NLO ones, 
$\mu_\chi=4 \pi F$,  $M^2=2 \Sigma m/F^2$ 
and $\delta=m_0^2/3 (4\pi F)^2$.
A comparison of Eq.~(\ref{eq:pappante}) with the data
in Fig.~\ref{fig:moltobella} indicates small
corrections due to quenched chiral logs and/or higher order terms
in this range. 

Since GW fermions do not suffer
from exceptional configurations, lighter pion masses 
can be simulated, provided the physical volume and the cut-off are large enough.
First exploratory  studies 
in the region $200\lesssim M_P \lesssim 400$ MeV have been reported at 
this conference \cite{Dong:2001fm,Hasenfratz:2002rp,Gattringer:2002sb}. 
\begin{figure}[htb]
\vspace{15pt}
\begin{center}
\epsfig{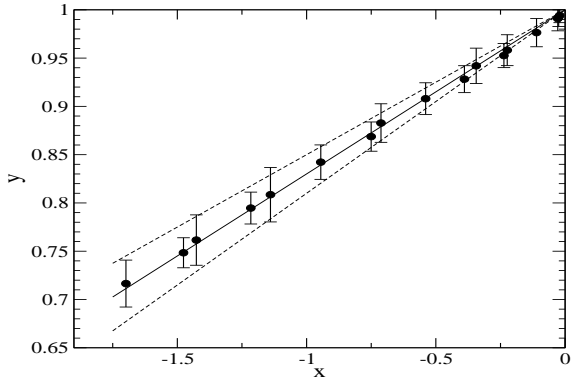}
\vspace{-1.0cm}
\caption{$y$ vs $x$ with FP action from Ref.~\cite{Gattringer:2002sb}.}
\label{fig:largenenough}
\end{center}
\end{figure}

The presence of  quenched chiral logs can be tested in
the double ratio of meson masses with non-degenerate quarks 
\cite{Bernard:1992mk,Aoki:2002fd}
\be
y=\frac{4 m_1 m_2}{(m_1+m_2)^2} 
  \frac{M^2_{P,12}}{M^2_{P,11}} 
  \frac{M^2_{P,12}}{M^2_{P,22}}\; .
\ee 
At NLO 
\be\label{eq:yexp}
y = 1 + \delta x + \frac{\alpha}{3 (4\pi F)^2}\frac{2\Sigma}{F^2}\, w + O(m_1^2,m_2^2)\; ,
\ee
where 
\ba
x & = & 2 + \frac{m_1+m_2}{m_1-m_2}\log{\Bigl(\frac{m_2}{m_1}\Bigr)}\\
w & = & \Bigl(\frac{2 m_1 m_2}{m_2-m_1}\log{\Bigl(\frac{m_2}{m_1}\Bigr)} - m_1 -m_2 \Bigr)
\ea
and $\mu_\chi$ appears explicitly in Eq.~(\ref{eq:yexp}) at higher orders
only.

Using the fixed-point (FP) 
and chirally improved (CI) actions, the Bern--Graz--Regensburg (BGR) collaboration 
computed meson masses with non-degenerate
quarks~\cite{Hasenfratz:2002rp,Gattringer:2002sb}.
In Fig.~\ref{fig:largenenough} data obtained with the FP operator 
on a lattice with $a \simeq 0.15$~fm and $L\simeq 2.4$~fm are shown.
By neglecting $2\alpha\Sigma w/3 F^2(4\pi F)^2 $ and higher order terms in 
Eq.~(\ref{eq:yexp}), they obtain $\delta=0.17(2)$ and $\delta=0.18(2)$
for the FP and CI actions respectively \cite{Gattringer:2002sb}. 
More studies are needed to properly assess the systematics due to finite volume effects
and/or finite lattice spacing, and to remove the uncertainties due to the leftover 
explicit symmetry breaking of these actions. An extensive comparison
of these results with those obtained in the past with standard actions 
can be found in \cite{HH}.

\section{The chiral condensate}
\begin{figure}[htb]
\vspace{15pt}
\hspace{-0.5cm}
%\begin{center}
\epsfig{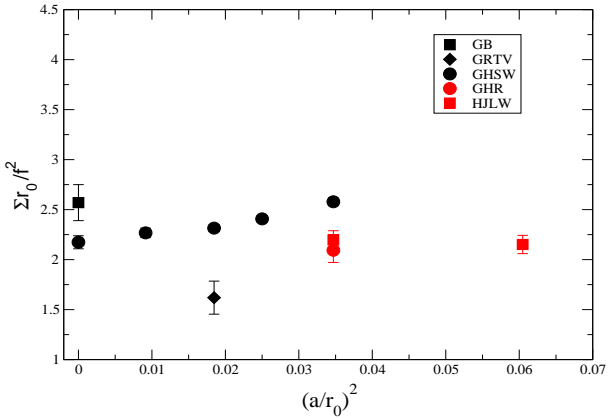}
\caption{Chiral condensate vs $a$: black squares \cite{Gupta:1996sa}, 
black diamonds \cite{Giusti:1998wy}, black circles \cite{Garden:1999fg}, 
red circles \cite{Giusti:2001pk}, red squares \cite{Hernandez:2001yn}.}
\label{fig:cond_all}
%\end{center}
\end{figure}

The chiral condensate $\Sigma$ has been computed with Wilson-type
fermions by fitting $M_P^2/2m$ with the LO term on the r.h.s.
of Eq.~(\ref{eq:pappante}) \cite{Gupta:1996sa,Giusti:1998wy}.
The same analysis has been repeated with overlap fermions in 
Refs. \cite{Giusti:2001pk,Hernandez:2001yn} and the results are reported 
in Fig.~\ref{fig:cond_all} as well. Within the statistical errors,
the results are compatible with the continuum extrapolated 
value obtained with non-perturbatively improved Wilson fermions \cite{Garden:1999fg}.
This represents a further indication that $O(a^2)$ effects are moderate 
with overlap fermions. More studies with lighter quark masses and larger 
volumes are needed to properly assess the systematics of these encouraging
results.
\begin{figure}[htb]
\vspace{15pt}
%\begin{center}
\hspace{-0.8cm}\epsfig{file=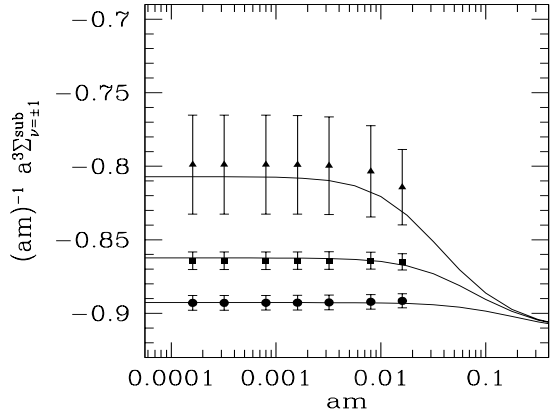,width=8.0cm,height=7.0cm}
\vspace{-1.5cm}
\caption{Quark mass dependence of the scalar condensate for three
volumes: $8^4$(circles), $10^4$(squares), and $12^4$(triangles), from 
Ref.~\cite{Hernandez:1999cu}.}
\label{fig:JLH}
%\end{center}
\end{figure}

Properties of QCD Green's functions 
in a finite box of linear extension $L$ 
and with very light quark masses 
can be studied within chiral perturbation theory \cite{Gasser:1987ah}.
If $2m\Sigma L^4 \sim 1$, $L\gg 1/(4\pi F)$ and $p^2\sim 1/L^2$,
the correlation functions can be expanded in powers of a parameter 
$\epsilon$ with
\be
\frac{\sqrt{2\Sigma m }}{\Lambda^2_\chi} \sim \frac{p^2}{\Lambda_\chi^2}
\sim  O(\epsilon^2)\; 
\ee
and $\Lambda_\chi$ being the cut-off of the 
effective theory \cite{Gasser:1987ah}. At leading order
the partition function is given by
\ba
\hspace{-0.4cm}& & Z(m,\theta) =  \int_{{\rm SU}(N_f)} 
d U_0 d \xi \times \\
& & 
\exp{\Bigl[}\frac{1}{2}\int d^4 x \Tr (\partial_\mu \xi \partial_\mu \xi)
+   z \mbox{\rm Re}\Tr({\rm e}^{i \theta/N_f} U_0)\Bigr]\nonumber
\ea
where the pion field is factorized as $U(x)=U_0\exp{(i\sqrt{2}\xi(x)/F)}$ 
and $z=m \Sigma V$. The integral over the global mode
\be
\int_{{\rm SU}(N_f)} \hspace{-0.8cm} d U_0 \exp{
\Bigl[z \mbox{\rm Re}\Tr( {\rm e}^{i \theta/N_f} U_0)\Bigr]}
\ee
needs to be done exactly, while a reordered chiral 
perturbation theory applies to the non-zero integration mode 
$\xi(x)$. Partition functions $Z_\nu(m)$ and correlations can be defined
in sectors of fixed topology 
by Fourier transforming in $\theta$ \cite{Leutwyler:1992yt}.
By comparing the chiral perturbation 
theory expectations for the correlation functions with the lattice data, the basic 
assumption of spontaneous symmetry breaking can in principle be verified and 
the low energy constants (LEC) of the 
chiral Lagrangian extracted. Properties of QCD in the infinite volume 
limit can then be recovered. 

The $\epsilon$-expansion has been extended to 
quenched QCD in Refs.~\cite{Osborn:1998qb,Damgaard:1998xy}.
At the leading order, the chiral condensate in a sector of 
fixed topology $\nu$ reads
\be\label{eq:cond_lead}
\frac{\Sigma_\nu (z)}{\Sigma} = z \Bigl( I_\nu(z) K_\nu(z)
+ I_{\nu + 1}(z) K_{\nu-1}(z)\Bigr) + \frac{\nu}{z}
\ee
where $I_\nu(z)$ and $K_\nu(z)$ are modified Bessel 
functions \cite{Damgaard:1998xy}. One-loop corrections 
give~\cite{Damgaard:2001xr,Damgaard:2001js}
\be\label{eq:condNLO}
\Sigma_\nu^{\rm 1-loop}(z) = \frac{z'}{z} \Sigma_\nu(z') 
\ee
where $z'= m\Sigma_{\rm eff} V$,
\ba
\Sigma_{\rm eff}(V) & = & \Sigma
\Bigl[ 
1 - \frac{m_0^2}{3 (4\pi F)^2}\Bigl(\tilde \beta_2 + 
\log{\Bigl(\frac{L_0^2}{L^2}\Bigr)}\Bigr)
\nonumber\\
& - & \frac{\alpha}{3 (4\pi F L)^2}
\tilde \beta_1\Bigr]\; ,
\ea
$1/L_0$ is the renormalization scale and 
$\tilde \beta_i$ are two universal 
``shape coefficients'' \cite{Damgaard:2001js}.
It is interesting to note that the expansion parameter 
$\epsilon \sim 1/(4\pi L F)$ in 
Eq.~(\ref{eq:condNLO}) is comparable to 
$M_P^2/(4\pi F)^2$ in Eq.~(\ref{eq:pappante})
for light pions ($M_P\sim 1/L$), while the prefactors
turn out to be different.

Quenched QCD in the $\epsilon$-regime has been explored 
on the lattice in Refs.~\cite{Edwards:1998wx,Hernandez:1999cu}. 
In Fig.~\ref{fig:JLH} 
an example of the 
results obtained  in Ref.~\cite{Hernandez:1999cu}  with overlap fermions 
for the vacuum expectation 
value of the scalar density in the topological sector 
$\nu=\pm 1$ at different volumes and various masses is shown. 
Even with poor statistics and small volumes,
a signal compatible with the expectations of 
chiral perturbation theory has been reported. Analogous
analyses have been attempted in 
Ref.~\cite{DeGrand:2001ie,Hasenfratz:2002rp}.
More studies at larger volumes and higher 
statistics are needed to confirm the indications of 
these  exploratory investigations. 

\section{Topological susceptibility}
In the chiral limit, the Fourier transform of the singlet axial WI 
in Eq.~(\ref{WTIREN}) for $\hat{\cal O}=\hat{Q}$ reads 
\ba\label{GQQNC}
\chi_t(p) & = & a^4 \sum_x \, \mbox{e}^{-ipx}\<\hat Q(x) \hat Q(0)\>+{\rm{CT}}(p) \\
& = & \frac{a^4}{2 N_f} \sum_x \, \mbox{e}^{-ipx}  
 \<\partial^*_\mu \hat A^0_\mu(x) \hat Q(0)\> +{\rm{CT}}(p)\nonumber
\ea 
where the same contact term ${\rm{CT}}(p)$ has been added to both sides of  
Eq.~(\ref{GQQNC}) to make them separately finite. 
${\rm{CT}}(p)$ is a fourth-degree polynomial, which can be chosen to vanish 
at $p=0$ since in this case the second line of Eq.~(\ref{GQQNC}) is certainly finite 
in the absence of zero-mass particles in the singlet channel. 
As a consequence, the topological 
susceptibility vanishes in the chiral limit, i.e. $\chi_t(0) = 0$ 
\cite{Giusti:2001xh}.

For $M_P\rightarrow 0$, the leading chiral behaviour
\be
\chi_t(0) = \frac{F^2 M_P^2}{2 N_f} + O(M_P^4)
\ee
is recovered by comparing the integrated singlet axial WIs with 
$\hat{\cal O}=\hat{Q}$ and $\hat{\cal O}=\hat{P}^0$ \cite{Chandrasekharan:1998wg}.  
\begin{figure}[htb]
\vspace{15pt}
%\begin{center}
\epsfig{file=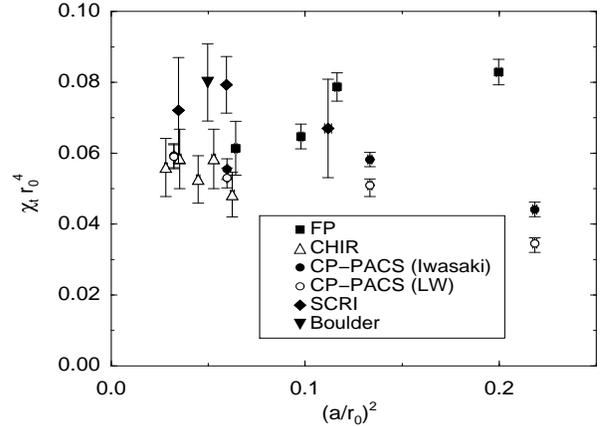,width=7.5cm,height=5.5cm}
\caption{Summary for the quenched topological susceptibility from \cite{Hasenfratz:2002rp}:
filled boxes \cite{Hasenfratz:2002rp}, empty triangles \cite{Gattringer:2002mr}, 
filled circles \cite{AliKhan:2001ym}, empty circles \cite{AliKhan:2001ym}, 
filled diamonds \cite{Edwards:1998sh}, filled triangles \cite{DeGrand:2002gm}.}
\label{fig:q2}
%\end{center}
\end{figure}

In the chiral limit, $\chi_{t}(p)$ satisfies a three-times-subtracted 
dispersion relation 
\ba
\chi_{t}(p) & = & b_1 + b_2 p^2 + b_3 (p^2)^2 \\ 
& - & \frac{R^2_{\eta'}}{p^2 + M^2_{\eta'}} + (p^2)^3
\int_{\rm cut} \frac{\rho(t)}{(t+p^2)t^3} dt\; ,\nonumber
\ea
where the contribution of the $\eta'$ meson has been 
separated since it is expected to dominate the dispersive 
integral \cite{Seiler:1987ig}.
For $p^2\rightarrow 0$, the ``sum rule'' $\chi_{t}(0) = 0$ implies
\be
b_1 = \frac{F^2_{\eta'} M^2_{\eta'}}{2 N_f} \; ,
\ee
where $R^2_{\eta'} = F^2_{\eta'} M^4_{\eta'}/2N_f$. Under the 
``smooth-quenching hypothesis'',
the Witten--Veneziano formula is obtained \cite{Witten:1979vv,Veneziano:1979ec}
\ba\label{eq:WV}
\displaystyle{
\frac{F^2 M_{\eta'}^2}{2N_f} \Bigr{|}_{\frac{N_f}{N_c}=0}}& = & 
a^4 \displaystyle{\sum_x \,\langle Q(x) Q(0)\rangle_{{\rm YM}}} 
\nonumber\\
& = & \frac{\langle(n_{-}-n_{+})^2\rangle}{V}\; ,
\ea
where the quantum average has to be done in the pure Yang--Mills (YM) 
theory \cite{Giusti:2001xh}. 

By using Wilson and staggered fermions it was argued 
that \cite{Bochicchio:1984hi,Smit:1986fn}
\be
\chi_{t}(0)
= \lim_{m\rightarrow 0}
\Bigl(\frac{2 m}{2 N_f}\Bigr)^2 a^4 \sum_x \langle P^0(x) P^0(0) 
\rangle^{{\rm ZV}}_{{\rm Quen}}
\ee
where only Zweig-violating (ZV) diagrams are included.
With GW fermions, this formula is the algebraic equivalent 
of Eq.~(\ref{eq:WV}) \cite{Chandrasekharan:1998wg}.

The topological susceptibility defined by using Neuberger's operator 
has been computed for 
several lattice spacings and volumes for a pure SU$(3)$ YM theory ~\cite{Edwards:1998sh}. 
Analogous computations have been performed in the past year with overlap 
fermions \cite{DeGrand:2002gm}, FP and CI actions \cite{Hasenfratz:2002rp,Gattringer:2002mr}. 
A summary of the results obtained can be found in \cite{Hasenfratz:2002rp} and is 
reported in Fig.~\ref{fig:q2}. The central value of $\chi_t$ 
obtained with GW fermions is 
quite stable as a function of the lattice spacing  
and is also compatible with the value obtained by other 
approaches \cite{Boyd:1997nt,Teper:1999wp}. 
More work is needed at larger volumes and smaller lattice 
spacings before these encouraging indications are fully confirmed 
and the magnitude of the systematic error is properly assessed.

Exploratory computations of $\chi_t$ for $N_c> 3$ have been performed
in the past year \cite{Lucini:2001ej,Cundy:2002hv,DelDebbio:2002xa}.
The results are compatible with a smooth large-$N_c$ limit and a 
non-zero $\chi_t$ in the limit $N_c\rightarrow \infty$.

\section{K$\rightarrow \pi\pi$ decays}
Non-leptonic $K\rightarrow \pi\pi$ amplitudes can be parametrized as 
\ba 
T[K^+\rightarrow\pi^+\pi^0] & = & \sqrt{3\over 2} A_2 e^{i\delta_2}\\
T[K^0\rightarrow\pi^+\pi^-] & = & \sqrt{2\over 3} A_0 e^{i\delta_0}+ \sqrt{1\over
3} A_2 e^{i\delta_2}\nonumber\\
T[K^0\rightarrow\pi^0\pi^0] & = & \sqrt{2\over 3} A_0 e^{i\delta_0}-2\sqrt{1\over
3} A_2 e^{i\delta_2}\nonumber
\ea
where $\delta_{I}$ and $A_I$ are the $\pi\pi$ phase shifts and 
the isospin amplitudes for $I=0,2$. Direct and indirect CP violation are parametrized by  
\be
\varepsilon' = \frac{1}{\sqrt{2}} {\rm e}^{i\Phi}
\frac{{\rm Re}A_{2}}{{\rm Re}A_{0}}
\left(\frac{{\rm Im}A_{2}}{{\rm Re}A_{2}}-\frac{{\rm Im}A_{0}} {{\rm Re}A_{0}}\right)
\ee
and 
\be
\varepsilon = \frac{T[K_L\rightarrow (\pi\pi)_{0}]}
                  {T[K_S\rightarrow (\pi\pi)_{0}]} 
\ee
respectively. Experimental results reveal $\Phi=\pi/2 + \delta_2 - \delta_0\thickapprox \pi/4$,
a $\Delta I=1/2$ selection rule $|A_0/A_2| \simeq  22.2$
and the presence of  
direct and indirect CP violation in nature: 
\ba
\mbox{Re}(\varepsilon'/\varepsilon) & = &  (16.6 \pm 1.6)\times 10^{-4} \;\; \;\;\;\;\;\cite{ichep2002} \nonumber\\
 | \varepsilon |\;\;\; & = & (2.282 \pm 0.017)\times 10^{-3}\;\; \cite{pdg2002}\; .
\ea
Phenomenological analyses of the unitarity 
triangle indicate that the Standard Model picture
of indirect CP violation in the kaon system is consistent with that of 
B decays and oscillations \cite{Ciuchini:2000de,Hocker:2001xe}. 

The $\Delta I=1/2$ rule and the 
value of $\varepsilon^\prime/\varepsilon$ can be explained 
within the Standard Model only if the strong interactions crucially affect 
these non-leptonic weak transitions 
(see \cite{Ciuchini:1999xi,Bertolini:2002rj,deRafael} for recent reviews). 
In this case a more complicated blend of ultraviolet and infrared effects 
prevented reliable determinations of the relevant matrix elements. 
Power-divergent subtractions can be needed
to construct the renormalized operators that enter 
the effective Hamiltonian \cite{Bernard:wf,Maiani:1986db,Dawson:1997ic,Capitani:2000bm}. 
In the infrared, the continuation 
of the theory to Euclidean space-time and the use of finite volumes in 
numerical simulations generate a non-simple relation between the physical 
amplitudes and those computed on the lattice \cite{Maiani:ca,Lellouch:2000pv,Lin:2001ek}. 

\subsection{The $\Delta I=1/2$ rule}
By using the operator product expansion (OPE), the CP-conserving 
$\Delta S =1$ effective Hamiltonian above the charm threshold is given by 
\[
H_{\rm eff}^{\Delta S =1} =  
\frac {G_F} {\sqrt{2}}
\Bigl[ C_{+}(\mu) \widehat {\cal O}_{+}(\mu) + 
       C_{-}(\mu) \widehat {\cal O}_{-}(\mu) \Bigr]\; ,
\]
where the Wilson coefficients $C_{\pm}(\mu)$ 
are known at the NLO \cite{Buras:1993dy,Ciuchini:1993vr}
and the bare operators are 
\ba
{\cal O}_{\pm} & = & \Bigl[({\bar s}^{a} \gamma_\mu P_- \tilde u^{b} )
        ({\bar u}^{b} \gamma_\mu P_- \tilde d^{a}) \\
& \pm &  
({\bar s} \gamma_\mu P_- \tilde u)
        ({\bar u} \gamma_\mu P_- \tilde d)\Bigr]
                -  (u \rightarrow c)\; .\nonumber
\ea
The contributions that arise when the top quark is integrated out
are heavily suppressed by CKM factors  
and can be neglected. ${\cal O}_{\pm}$ belong 
to different chiral multiplets and are CPS-even. 
In correlation functions at non-zero physical distance, ${\cal O}_{\pm}$ cannot
mix between themselves or with other four-fermion operators,  
but only with the dimension-six operator \cite{Capitani:2000bm,kpipi_club}
\be
{\cal Q}_m = (m^2_u -m^2_c)\Bigl[m_d (\bar s P_+ \tilde d) + m_s (\bar s P_- \tilde d)\Bigr]\; .
\ee
The renormalized operators are
\be
\widehat {\cal O}_{\pm}(\mu) =  Z_{\pm}(\mu) 
\Bigl[ {\cal O}_{\pm} + b_{\pm}^m {\cal Q}_{m}\Bigr]\; , \;\;
\ee
where $Z_{\pm}(\mu)$ are logarithmic-divergent 
renormalization constants and $b_{\pm}^m$ are 
suppressed by a factor  $\alpha_s$.
{\it No power-divergent subtractions are needed}
to renormalize ${\cal O}_{\pm}$ when fermions with an exact 
chiral symmetry are used \cite{Capitani:2000bm}.

For $m_s \neq m_d$,
\ba
{\cal Q}_m & = & (m^2_u -m^2_c) 
\partial^*_\mu\Bigl[\frac{m_d+m_s}{m_s-m_d}\, {\cal V}_\mu^{sd}\\
& + & 
\frac{m_d-m_s}{m_s+m_d}\, {\cal A}_\mu^{sd} \Bigr]\nonumber
\ea
and it does not contribute to matrix elements which 
preserve four-momentum \cite{Bernard:wf,Maiani:1986db}.

If the charm is integrated out not only potentially large
contributions of $O(\mu^2/m_c^2)$ are neglected,
but ultraviolet power divergences can arise
in the renormalization pattern of the relevant 
four-fermion operators. In this case the 
$\Delta S =1 $ effective Hamiltonian can be written as
\be\label{eq:nocharm}
H_{\rm eff}^{\Delta S =1} =  \displaystyle \frac {G_F} {\sqrt{2}} 
\sum_{i=1}^{10} C_i(\mu) \widehat {\cal Q}_i(\mu) \; .
\ee 
The so-called QCD-penguin operators are 
\ba
{\cal Q}_{3,5} &=& ({\bar s} \gamma_\mu P_- \tilde d)
    \sum_{q=u,d,s}({\bar q} \gamma_\mu P_\mp \tilde q) \\
{\cal Q}_{4,6} &=& ({\bar s}^{a} \gamma_\mu P_- \tilde d^{b})
    \sum_{q=u,d,s}({\bar q}^{b} \gamma_\mu P_\mp \tilde q^{a})
\ea
(see Refs.~\cite{Ciuchini:1999xi,Bertolini:2002rj}
for definitions of the other operators).
At non-zero physical distance, mixing with two lower-dimensional operators 
\ba
{\cal Q}_p & = & m_d (\bar s P_+ \tilde d) + m_s (\bar s P_- \tilde d)\\
{\cal Q}_\sigma & = & m_d (\bar s F_{\mu\nu} \sigma_{\mu\nu}P_+ \tilde d)
+ m_s (\bar s F_{\mu\nu} \sigma_{\mu\nu}P_- \tilde d)\nonumber
\ea
can occur and power-divergent subtractions are needed even 
with Ginsparg--Wilson fermions.

With Wilson fermions, 
only CPS and flavour symmetry can be used to determine the 
renormalization pattern of ${\cal O}^{\pm}$. Even with an active charm, 
a quadratic divergent contribution needs to 
be subtracted in the parity-conserving sector
\ba
\widehat {\cal O}^{\rm PC}_{\pm}(\mu) & = & {\cal Z}_{\pm}(\mu) 
\Bigl[{\cal O}^{\rm PC}_{\pm} + \sum_j {b}^j_\pm {\cal O}^{\pm}_j +\nonumber\\
& + &  {b}_{\pm}^{\tau}{\cal Q}_\tau + 
\frac{{b}_{\pm}^s}{a^2} {\cal Q}_s \Bigr]
\ea
where 
\ba
{\cal Q}_s     & = & (m_u - m_c) \bar s  d \\
{\cal Q}_\tau  & = & (m_u - m_c) \bar s\, \sigma_{\mu\nu} F_{\mu\nu} d  \; .
\ea
and ${\cal O}_j^\pm$ are four fermion operators with wrong chirality.
In this case only the flavour part of the GIM mechanism survives due 
to the explicit breaking of chiral symmetry \cite{Maiani:1986db}.

The LECs of the CP-conserving $\Delta S=1$ electroweak chiral 
Lagrangian can be extracted from three-point correlation functions,
thus avoiding the infrared problem that affects the direct 
computation of the $K\rightarrow \pi\pi$ matrix elements on the lattice 
\cite{Bernard:wf}. No large cancellations among leading order terms 
are expected in the ratio $|A_0/A_2|$; an enhancement
should therefore be visible already at this order. As a result, a combined 
use of fermions 
with an exact chiral symmetry and chiral perturbation theory can be 
the starting point to attack the $\Delta I=1/2$ rule.

In the past years the RBC and the CP-PACS collaborations have studied the 
$\Delta I=1/2$ rule with domain-wall fermions with a finite fifth dimension 
\cite{Blum:2001xb,Noaki:2001un}.  They have 
computed the $K\rightarrow \pi$ and $K\rightarrow 0$ matrix elements for 
the operators of the $\Delta S=1$ effective Hamiltonian in Eq.~(\ref{eq:nocharm})
and used LO chiral perturbation theory to recover 
the physical amplitudes. Although with large statistical and systematic
errors, both groups demonstrated that a controlled numerical signal can be 
obtained for these matrix elements. The systematics of these important results 
can be reduced by using fermions with an exact chiral symmetry, lighter quark 
masses and by considering the effective Hamiltonian with a dynamical charm.
\begin{figure*}[htb]
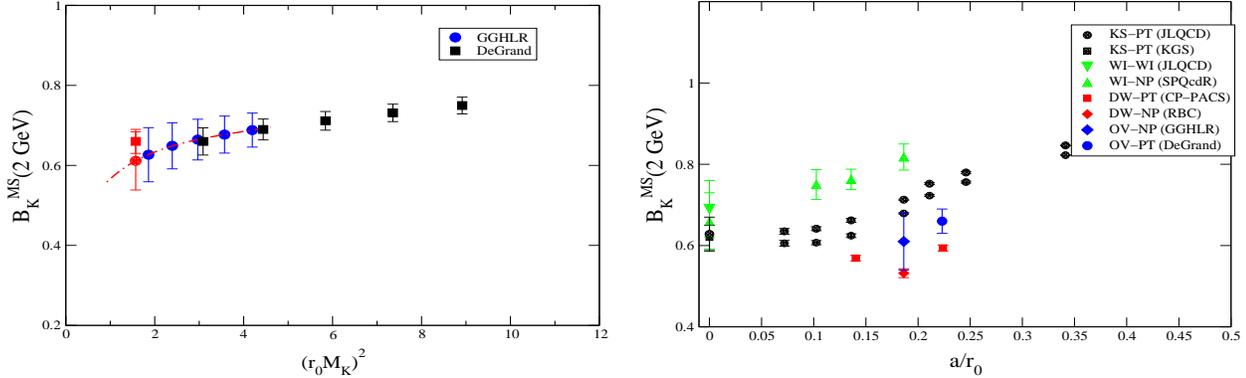

%\vspace{-1.0cm}
\catcode`?=\active \def?{\kern\digitwidth}
%\begin{center}
\hspace{-0.6cm}\begin{tabular}{cc}
\mbox{\epsfig{file=bk_all.eps,width=5.0cm,height=8.0cm,angle=270}} &
\mbox{\epsfig{file=bk_summary.eps,width=5.0cm,height=8.0cm,angle=270}} \\
\end{tabular}
%\vspace{-1.0cm}
\caption{On the left $B^{\msbar}_K(2\, {\rm GeV})$ vs $(r_0 M_K)^2$ for plain  
(circles) \cite{Marseille} and  NNC-HYP (squares) \cite{DeGrand} overlap. 
On the right, $B^{\msbar}_K(2\, {\rm GeV})$
vs the lattice spacing from various actions: black circles \cite{Aoki:1997nr},
black squares \cite{Kilcup:1997ye}, green triangles down \cite{Aoki:1999gw}, 
green triangles up \cite{Becirevic:2002mm}, 
red squares \cite{AliKhan:2001wr}, red diamonds \cite{Blum:2001xb}, 
blue diamonds \cite{Marseille}, blue circles \cite{DeGrand}.\label{fig:BK}}
%\end{center}
\end{figure*}

A different avenue is being followed in Ref.~\cite{kpipi_club}. It is 
conceivable that the LECs of the weak chiral Lagrangian can be extracted by 
studying the weak interactions in the $\epsilon$-regime \cite{kpipi_club}. 
A numerical feasibility study of this approach on the lattice is under way.
If feasible, it will be very interesting to compare the results of the 
LECs in the two regimes.

For direct CP violation, both the ultraviolet and the infrared problems are 
more severe. An active charm does not mitigate the ultraviolet 
renormalization, and divergent power subtractions are necessary.
The cancellation between two large competing contributions 
from ${\cal Q}_6$ and ${\cal Q}_8$ renders $\varepsilon'/\varepsilon$ very sensitive 
to higher order corrections in chiral 
perturbation theory. Leading-order terms may not 
be sufficient to reach a reliable prediction in the Standard 
Model \cite{Bertolini:1998vd,Pallante:2001he}. 

\subsection{$K^0$--$\bar{K}^0$ mixing: $\varepsilon$}
By using the OPE, the 
$\Delta S =2$ effective Hamiltonian is given by 
\be
H^{\Delta S=2}_{\rm eff} = \frac{G^2_F M^2_W}{4\pi^2}
 {\cal C}_1(\mu) \, \widehat{ \cal O}_1 (\mu) + {\rm h. c.}\; ,
\ee
where the expression of the Wilson coefficient $ {\cal C}_1(\mu)$ 
is known at NLO \cite{Herrlich:1995hh} and the corresponding
bare four-fermion operator 
\be
{\cal O}_{1} = (\bar s \gamma_\mu P_- \tilde d) (\bar s \gamma_\mu P_- \tilde d )
\ee
is multiplicatively renormalizable.
The matrix element that encodes the long-distance QCD contributions 
to $\varepsilon$ 
\be
\langle \bar K^0| \widehat {\cal O}_1(\mu) |K^0\rangle \equiv
\frac{4}{3}  F_K^2 M_K^2 \widehat B_K(\mu)
\ee
has been computed with overlap fermions in the last year \cite{Marseille,DeGrand}. 
A plain overlap
action has been used in Ref.~\cite{Marseille} for a lattice of linear extension 
$L\simeq 1.5$~fm, with a spacing $a\simeq0.093$~fm
and for degenerate light quark masses in the range 
$m_s/2\lesssim m \lesssim m_s$. The 
RI/MOM non-perturbative renormalization procedure
has been implemented to compute the logarithmic divergent 
renormalization constant. NNC-HYP overlap fermions 
have been used in Ref.~\cite{DeGrand} for a lattice of linear extension 
$L\simeq 1.5$~fm, with a spacing $a\simeq0.125$~fm
and for degenerate light quark masses in the range 
$m_s\lesssim m \lesssim 2.5 m_s$. The operator 
has been renormalized using one-loop 
perturbation theory. The results of the two computations are 
in very good agreement in the 
common range of simulated masses, as 
shown in Fig.~\ref{fig:BK}.
 
In Ref.~\cite{Marseille} the results have been slightly
extrapolated to the physical point  
by using the functional form 
\ba
\widehat B^{\msbar}_K(2\, \mbox{GeV}) & = & B_0\Bigl( 1-3 \Bigl(\frac{M_K}{4 \pi F}\Bigr)^2
\log \Bigl(\frac{M_K^2}{\Lambda_\chi^2}\Bigr)\nonumber\\ 
& + &  b\, \Bigl( \frac{M_K}{4\pi F}\Bigr)^4\Bigr)\;
\ea
with $F=F_\pi^{\rm phys}$, while a linear extrapolation has been performed in \cite{DeGrand}.
Including the statistical errors only, the preliminary results
\ba
\widehat B^{\msbar}_K(2\, \mbox{GeV}) & = & 0.61 \pm 0.07 \qquad \mbox{\cite{Marseille}}\\
                             & = & 0.66 \pm 0.04 \qquad \mbox{\cite{DeGrand}}
\ea
are in very good agreement. More studies are needed to properly assess the magnitude of the 
various systematic errors. 

A comparison with other determinations obtained 
with different regularizations is shown in the second plot of Fig.~\ref{fig:BK}.
Even if the statistical errors are large, the agreement
with the continuum-limit world averages based on staggered results
in Ref.~\cite{Aoki:1997nr} is very good. Results obtained with domain-wall fermions
with a finite fifth dimension are below the overlap determinations, but 
still compatible within errors \cite{Blum:2001xb,AliKhan:2001wr}.

\section{Conclusions}
Studying QCD with an exact chiral symmetry at finite lattice spacing 
implies many theoretical advantages: in the ultraviolet it simplifies the subtraction 
of divergences in composite operators,
in the infrared it allows one to simulate massless quarks
and it provides a natural definition for the topological charge and the 
topological susceptibility.

Large-scale QCD simulations with fermions with exact chiral symmetry are feasible 
with known algorithms and the present generation of computers, 
at least in the quenched approximation. A regime of light quark masses not accessible 
with standard fermions has already been reached. 

First results for the meson
spectrum, light quark masses, the chiral condensate and $B_K$ 
are in good agreement with previous determinations and suggest moderate 
discretization errors for Neuberger's fermions.

Important long-standing problems 
such as the $\Delta I=1/2$ rule in $K\rightarrow\pi\pi$ decays are greatly 
simplified and can be attacked. A combined use of chiral perturbation theory and 
fermions with exact chiral symmetry may lead to important new 
phenomenological informations in the next few years.

\section*{Acknowledgements}
I warmly thank S.~Capitani, P.~Hern\'andez, C.~Hoelbling, 
G.~Isidori, L.~Lellouch, V.~Lubicz, M.~L\"uscher, G.~Martinelli,
H.~Neuberger, C.~Rebbi, G.C.~Rossi and M.~Testa 
for interesting and stimulating discussions.
It is a pleasure to thank T.W.~Chiu, T.~DeGrand, K.F.~Liu,
S.~Necco, M.~Papinutto, A.~Yamaguchi and the members of the
BGR and RBC collaborations for sending their results and
for interesting discussions about their work.
Many thanks to the organizers for the very stimulating atmosphere 
of the conference.


\begin{thebibliography}{9}
\bibitem{Rubakov:bb}
V.~A.~Rubakov and M.~E.~Shaposhnikov,
Phys. Lett. B125 (1983) 136.
%%CITATION = PHLTA,B125,136;%%
\bibitem{Callan:sa}
C.~G.~Callan and J.~A.~Harvey,
Nucl. Phys. B250 (1985) 427.
%%CITATION = NUPHA,B250,427;%%
\bibitem{Kaplan:1992bt}
D.~B.~Kaplan, Phys. Lett. B288 (1992) 342. %[arXiv:hep-lat/9206013].
%%CITATION = HEP-LAT 9206013;%%
\bibitem{NaraNeub}
R.~Narayanan and H.~Neuberger,
Phys. Lett. B302 (1993) 62; %[arXiv:hep-lat/9212019].
%%CITATION = HEP-LAT 9212019;%%
Phys. Rev. Lett.  71 (1993) 3251; %[arXiv:hep-lat/9308011].
%%CITATION = HEP-LAT 9308011;%%
Nucl. Phys. B412 (1994) 574 %[arXiv:hep-lat/9307006].
%%CITATION = HEP-LAT 9307006;%%
and B443 (1995) 305. %[arXiv:hep-lat/9411108].
%%CITATION = HEP-TH 9411108;%%
\bibitem{Neuberger}
H.~Neuberger, 
Phys. Lett. B417 (1998) 141; %[arXiv:hep-lat/9707022].
%%CITATION = HEP-LAT 9707022;%%
Phys. Rev. D57 (1998) 5417. %[arXiv:hep-lat/9710089].
%%CITATION = HEP-LAT 9710089;%%
\bibitem{Hernandez:1998et}
P.~Hern\'andez, K.~Jansen and M.~L\"uscher,
Nucl. Phys. B552 (1999) 363. %[arXiv:hep-lat/9808010].
%%CITATION = HEP-LAT 9808010;%%
\bibitem{Ginsparg:1982bj}
P.~H.~Ginsparg and K.~G.~Wilson,
Phys. Rev. D25 (1982) 2649.
%%CITATION = PHRVA,D25,2649;%%
\bibitem{Luscher:1998pq}
M.~L\"uscher,
Phys. Lett. B428 (1998) 342. %[arXiv:hep-lat/9802011].
%%CITATION = HEP-LAT 9802011;%%
\bibitem{PerfectA}
P.~Hasenfratz and F.~Niedermayer,
Nucl. Phys. B414 (1994) 785; %[arXiv:hep-lat/9308004].
%%CITATION = HEP-LAT 9308004;%%
T.~DeGrand et al.,
Nucl. Phys. B454 (1995) 587. %[arXiv:hep-lat/9506030].
%%CITATION = HEP-LAT 9506030;%%
\bibitem{Hasenfratz:1997ft}
P.~Hasenfratz,
Nucl. Phys. B (Proc. Suppl.) 63 (1998) 53. %[arXiv:hep-lat/9709110].
%%CITATION = HEP-LAT 9709110;%%
\bibitem{gattringer}
C. Gattringer, these proceedings.
\bibitem{Shamir:1993zy}
Y.~Shamir,
Nucl. Phys. B406 (1993) 90. %[arXiv:hep-lat/9303005].
%%CITATION = HEP-LAT 9303005;%%
V.~Furman and Y.~Shamir,
Nucl. Phys. B439 (1995) 54. %[arXiv:hep-lat/9405004].
%%CITATION = HEP-LAT 9405004;%%
\bibitem{Neuberger:1998wv} 
H.~Neuberger,
Phys. Lett. B427 (1998) 353. %[arXiv:hep-lat/9801031].
%%CITATION = HEP-LAT 9801031;%%
\bibitem{Fujikawa:1979ay}
K.~Fujikawa,
Phys. Rev. Lett. 42 (1979) 1195; 
%%CITATION = PRLTA,42,1195;%%
Phys. Rev. D21 (1980) 2848 [Erratum, ibid. D22 (1980) 1499].
%%CITATION = PHRVA,D21,2848;%%
\bibitem{Hasenfratz:1998ri}
P.~Hasenfratz, V.~Laliena and F.~Niedermayer,
Phys. Lett. B427 (1998) 125. %[arXiv:hep-lat/9801021].
%%CITATION = HEP-LAT 9801021;%%
\bibitem{kpipi_club}
L.~Giusti, P.~Hern\'andez, C.~Hoelbling, 
K.~Jansen, M.~Laine, L.~Lellouch,
M.~L\"uscher, P.~Weisz and H.~Wittig, in preparation.
\bibitem{Fujikawa:2002vj}
K.~Fujikawa, M.~Ishibashi and H.~Suzuki,
JHEP 0204 (2002) 046. %[arXiv:hep-lat/0203016].
%%CITATION = HEP-LAT 0203016;%%
\bibitem{Kikukawa:1998py}
Y.~Kikukawa and A.~Yamada,
Nucl. Phys. B547 (1999) 413. %[arXiv:hep-lat/9808026].
%%CITATION = HEP-LAT 9808026;%%
\bibitem{Giusti:2001xh}
L.~Giusti et al.,
Nucl. Phys. B628 (2002) 234. %[arXiv:hep-lat/0108009].
%%CITATION = HEP-LAT 0108009;%%
\bibitem{Aoki:1997xg}
S.~Aoki and Y.~Taniguchi,
%``One loop calculation in lattice QCD with domain-wall quarks,''
Phys. Rev. D59 (1999) 054510. %[arXiv:hep-lat/9711004].
%%CITATION = HEP-LAT 9711004;%%
\bibitem{Aoki:1998vv}
S.~Aoki et al.,
%S.~Aoki, T.~Izubuchi, Y.~Kuramashi and Y.~Taniguchi,
%``Perturbative renormalization factors of quark bilinear operators for  domain-wall QCD,''
Phys. Rev. D59 (1999) 094505. %[arXiv:hep-lat/9810020].
%%CITATION = HEP-LAT 9810020;%%
\bibitem{Aoki:1998hi}
S.~Aoki and Y.~Taniguchi,
%``One loop renormalization for the axial Ward-Takahashi identity in  domain-wall QCD,''
Phys. Rev. D59 (1999) 094506. %[arXiv:hep-lat/9811007].
%%CITATION = HEP-LAT 9811007;%%
\bibitem{Aoki:1999ky}
S.~Aoki et al.,
%S.~Aoki, T.~Izubuchi, Y.~Kuramashi and Y.~Taniguchi,
%``Perturbative renormalization factors of three- and four-quark  operators for domain-wall QCD,''
Phys. Rev. D60 (1999) 114504. %[arXiv:hep-lat/9902008].
%%CITATION = HEP-LAT 9902008;%%
\bibitem{Aoki:2000ee}
S.~Aoki and Y.~Kuramashi,
%``Perturbative renormalization factors of Delta(S) = 1 four-quark  
% operators for domain-wall QCD,''
Phys. Rev. D63 (2001) 054504. %[arXiv:hep-lat/0007024].
%%CITATION = HEP-LAT 0007024;%%
\bibitem{Aoki:2002iq}
S.~Aoki et al.,
%S.~Aoki, T.~Izubuchi, Y.~Kuramashi and Y.~Taniguchi,
%``Perturbative renormalization factors in domain-wall QCD with improved  gauge actions,''
arXiv:hep-lat/0206013.
%%CITATION = HEP-LAT 0206013;%%
\bibitem{Ishibashi:1999ik}
M.~Ishibashi et al.,
%``One-loop analyses of lattice QCD with the overlap Dirac operator,''
Nucl. Phys. B576 (2000) 501. %[arXiv:hep-lat/9911037].
%%CITATION = HEP-LAT 9911037;%%
%\cite{Alexandrou:1999wr}
\bibitem{Alexandrou:1999wr}
C.~Alexandrou, H.~Panagopoulos and E.~Vicari,
%``Lambda-parameter of lattice QCD with the overlap-Dirac operator,''
Nucl. Phys. B571 (2000) 257. %[arXiv:hep-lat/9909158].
%%CITATION = HEP-LAT 9909158;%%
\bibitem{Alexandrou:2000kj}
C.~Alexandrou et al.,
Nucl. Phys. B580 (2000) 394. %[arXiv:hep-lat/0002010].
%%CITATION = HEP-LAT 0002010;%%
\bibitem{Capitani:2000aq}
S.~Capitani,
Nucl.  Phys.  B592 (2001) 183 and %[arXiv:hep-lat/0005008].
%%CITATION = HEP-LAT 0005008;%%
B597 (2001) 313. %[arXiv:hep-lat/0009018].
%%CITATION = HEP-LAT 0009018;%%
\bibitem{Capitani:2000da}
S.~Capitani and L.~Giusti,
%``Perturbative renormalization of weak-Hamiltonian four-fermion operators  with overlap fermions,''
Phys. Rev. D62 (2000) 114506. %[arXiv:hep-lat/0007011].
%%CITATION = HEP-LAT 0007011;%%
\bibitem{Capitani:2000bm}
S.~Capitani and L.~Giusti,
Phys. Rev. D64 (2001) 014506. %[arXiv:hep-lat/0011070].
%%CITATION = HEP-LAT 0011070;%%
\bibitem{DeGrand:2002va}
T.~DeGrand,
%``One loop matching coefficients for a variant overlap action and some of  its simpler relatives,''
arXiv:hep-lat/0210028.
%%CITATION = HEP-LAT 0210028;%%
\bibitem{Blum:2001sr}
T.~Blum et al.,
%``Non-perturbative renormalisation of domain wall fermions: Quark  bilinears,''
Phys. Rev. D66 (2002) 014504. %[arXiv:hep-lat/0102005].
%%CITATION = HEP-LAT 0102005;%%
\bibitem{Giusti:2001pk}
L.~Giusti, C.~Hoelbling and C.~Rebbi,
Phys. Rev. D64 (2001) 114508 [Erratum, ibid. D65 (2002) 079903] 
%[arXiv:hep-lat/0108007].
%%CITATION = HEP-LAT 0108007;%%
and Nucl. Phys.  B (Proc. Suppl.) 106 \& 107 (2002) 739. %[arXiv:hep-lat/0110184].
%%CITATION = HEP-LAT 0110184;%%
\bibitem{Dong:2001fm}
S.~J.~Dong et al.,
%S.~J.~Dong, T.~Draper, I.~Horvath, F.~X.~Lee, K.~F.~Liu and J.~B.~Zhang,
%``Chiral properties of pseudoscalar mesons on a quenched 20**4 
%lattice  with overlap fermions,''
Phys. Rev. D65 (2002) 054507 and these proceedings. %[arXiv:hep-lat/0108020].
%%CITATION = HEP-LAT 0108020;%%
\bibitem{Hernandez:2001yn}
P.~Hern\'andez et al.,
%P.~Hernandez, K.~Jansen, L.~Lellouch and H.~Wittig,
JHEP 0107 (2001) 018; %[arXiv:hep-lat/0106011].
%%CITATION = HEP-LAT 0106011;%%
Nucl. Phys. B (Proc. Suppl.) 106 \& 107 (2002) 766. %[arXiv:hep-lat/0110199].
%%CITATION = HEP-LAT 0110199;%%
\bibitem{Blum:2001xb}
T.~Blum et al.  [RBC Collaboration],
%``Kaon matrix elements and CP-violation from quenched lattice QCD. I: The  3-flavor case,''
arXiv:hep-lat/0110075.
%%CITATION = HEP-LAT 0110075;%%
\bibitem{Marseille}
N. Garron et al., these proceedings.
\bibitem{Chiu:2002xm}
T.~W.~Chiu and T.~H.~Hsieh,
Phys. Rev. D66 (2002) 014506 and these proceedings. %[arXiv:hep-lat/0204009].
%%CITATION = HEP-LAT 0204009;%%
\bibitem{Hasenfratz:2002rp}
P.~Hasenfratz et al., 
%P.~Hasenfratz, S.~Hauswirth, T.~Jorg, F.~Niedermayer and K.~Holland,
arXiv:hep-lat/0205010.
%%CITATION = HEP-LAT 0205010;%%
\bibitem{Gattringer:2002sb} % BGR proceding lattice cumulativo
C.~Gattringer et al.  [BGR Collaboration],
these proceedings.
\bibitem{Allton:1996yv}
C.~R.~Allton et al.,
%C.~R.~Allton, V.~Gimenez, L.~Giusti and F.~Rapuano,
Nucl. Phys. B489 (1997) 427. %[arXiv:hep-lat/9611021].
%%CITATION = HEP-LAT 9611021;%%
\bibitem{Garden:1999fg}
J.~Garden et al.  [ALPHA Collaboration],
%J.~Garden, J.~Heitger, R.~Sommer and H.~Wittig  [ALPHA Collaboration],
Nucl. Phys. B571 (2000) 237. %[arXiv:hep-lat/9906013].
%%CITATION = HEP-LAT 9906013;%%
\bibitem{Aoki:2002fd}
S.~Aoki et al.  [CP-PACS Collaboration],
arXiv:hep-lat/0206009.
%%CITATION = HEP-LAT 0206009;%%
\bibitem{Colangelo:1997ch} % colangelo-pallante NLO
G.~Colangelo and E.~Pallante,
Nucl. Phys. B520 (1998) 433. %[arXiv:hep-lat/9708005].
%%CITATION = HEP-LAT 9708005;%%
\bibitem{Heitger:2000ay} % NLO di alpha
J.~Heitger et al. [ALPHA Collaboration],
%J.~Heitger, R.~Sommer and H.~Wittig  [ALPHA Collaboration],
Nucl. Phys. B588 (2000) 377. %[arXiv:hep-lat/0006026].
%%CITATION = HEP-LAT 0006026;%%
\bibitem{Bernard:1992mk}
C.~W.~Bernard and M.~F.~Golterman,
%``Chiral perturbation theory for the quenched approximation of QCD,''
Phys. Rev. D46 (1992) 853. %[arXiv:hep-lat/9204007].
%%CITATION = HEP-LAT 9204007;%%
\bibitem{Sharpe:1992ft}
S.~R.~Sharpe,
%``Quenched chiral logarithms,''
Phys.\ Rev. D46 (1992) 3146. %[arXiv:hep-lat/9205020].
%%CITATION = HEP-LAT 9205020;%%
\bibitem{Blum:2000kn}
T.~Blum et al.,
%``Quenched lattice QCD with domain wall fermions and the chiral limit,''
arXiv:hep-lat/0007038.
%%CITATION = HEP-LAT 0007038;%%
\bibitem{HH}
H. Wittig, these proceedings.
\bibitem{Gupta:1996sa}
R.~Gupta and T.~Bhattacharya,
Phys. Rev. D55 (1997) 7203. %[arXiv:hep-lat/9605039].
%%CITATION = HEP-LAT 9605039;%%
\bibitem{Giusti:1998wy}
L.~Giusti et al.,
%L.~Giusti, F.~Rapuano, M.~Talevi and A.~Vladikas,
%``The QCD chiral condensate from the lattice,''
Nucl. Phys. B538 (1999) 249. %[arXiv:hep-lat/9807014].
%%CITATION = HEP-LAT 9807014;%%
\bibitem{Hernandez:1999cu} 
P.~Hern\'andez, K.~Jansen and L.~Lellouch,
Phys. Lett. B469 (1999) 198. %[arXiv:hep-lat/9907022].
%%CITATION = HEP-LAT 9907022;%%
\bibitem{Gasser:1987ah}
J.~Gasser and H.~Leutwyler,
Phys. Lett. B188 (1987) 477.
%%CITATION = PHLTA,B188,477;%%
\bibitem{Leutwyler:1992yt}
H.~Leutwyler and A.~Smilga,
Phys. Rev. D46 (1992) 5607.
%%CITATION = PHRVA,D46,5607;%%
\bibitem{Osborn:1998qb}
J.~C.~Osborn, D.~Toublan and J.~J.~Verbaarschot,
Nucl. Phys. B540 (1999) 317. %[arXiv:hep-th/9806110].
%%CITATION = HEP-TH 9806110;%%
\bibitem{Damgaard:1998xy}
P.~H.~Damgaard et al.,
%P.~H.~Damgaard, J.~C.~Osborn, D.~Toublan and J.~J.~Verbaarschot,
Nucl. Phys. B547 (1999) 305. %[arXiv:hep-th/9811212].
%%CITATION = HEP-TH 9811212;%%
\bibitem{Damgaard:2001xr}
P.~H.~Damgaard,
Nucl. Phys. B608 (2001) 162. %[arXiv:hep-lat/0105010].
%%CITATION = HEP-LAT 0105010;%%
\bibitem{Damgaard:2001js}
P.~H.~Damgaard et al., 
%P.~H.~Damgaard, M.~C.~Diamantini, P.~Hernandez and K.~Jansen,
Nucl. Phys. B629 (2002) 445. %[arXiv:hep-lat/0112016].
%%CITATION = HEP-LAT 0112016;%%
\bibitem{Edwards:1998wx}
R.~G.~Edwards, U.~M.~Heller and R.~Narayanan,
%``A study of chiral symmetry in quenched QCD using the overlap-Dirac  operator,''
Phys. Rev. D59 (1999) 094510
[arXiv:hep-lat/9811030].
%%CITATION = HEP-LAT 9811030;%%
\bibitem{DeGrand:2001ie}
T.~DeGrand  [MILC Collaboration],
Phys. Rev. D64 (2001) 117501. %[arXiv:hep-lat/0107014].
%%CITATION = HEP-LAT 0107014;%%
\bibitem{Chandrasekharan:1998wg}
S.~Chandrasekharan,
Phys. Rev. D60 (1999) 074503. %[arXiv:hep-lat/9805015].
%%CITATION = HEP-LAT 9805015;%%
\bibitem{Seiler:1987ig}
E.~Seiler and I.~O.~Stamatescu,
MPI-PAE/PTh 10/87; E.~Seiler,
Phys. Lett. B525 (2002) 355. %[arXiv:hep-th/0111125].
%%CITATION = HEP-TH 0111125;%%
\bibitem{Witten:1979vv}
E.~Witten,
Nucl. Phys. B156 (1979) 269.
%%CITATION = NUPHA,B156,269;%%
\bibitem{Veneziano:1979ec}
G.~Veneziano,
Nucl. Phys. B159 (1979) 213.
%%CITATION = NUPHA,B159,213;%%
%\cite{Gattringer:2002mr}
\bibitem{Gattringer:2002mr}
C.~Gattringer, R.~Hoffmann and S.~Schaefer,
%``The topological susceptibility of SU(3) gauge theory near T(c),''
Phys. Lett. B535 (2002) 358. %[arXiv:hep-lat/0203013].
%%CITATION = HEP-LAT 0203013;%%
\bibitem{AliKhan:2001ym}
A.~Ali Khan et al.  [CP-PACS Collaboration],
%``Topological susceptibility in lattice QCD with two flavors of dynamical  quarks,''
Phys. Rev. D64 (2001) 114501. %[arXiv:hep-lat/0106010].
%%CITATION = HEP-LAT 0106010;%%
\bibitem{Edwards:1998sh}
R.~G.~Edwards, U.~M.~Heller and R.~Narayanan,
%``Spectral flow, chiral condensate and topology in lattice QCD,''
Nucl. Phys. B535 (1998) 403; %[arXiv:hep-lat/9802016].
%%CITATION = HEP-LAT 9802016;%%
%R.~G.~Edwards, U.~M.~Heller and R.~Narayanan,
%``Approach to the continuum limit of the quenched Hermitian Wilson-Dirac  operator,''
D60 (1999) 034502. %[arXiv:hep-lat/9901015].
%%CITATION = HEP-LAT 9901015;%%
\bibitem{DeGrand:2002gm} 
T.~DeGrand and U.~M.~Heller  [MILC collaboration],
Phys. Rev. D65 (2002) 114501. %[arXiv:hep-lat/0202001].
%%CITATION = HEP-LAT 0202001;%%
\bibitem{Bochicchio:1984hi}
M.~Bochicchio et al.,
%M.~Bochicchio, G.~C.~Rossi, M.~Testa and K.~Yoshida,
Phys.  Lett. B149 (1984) 487.
%%CITATION = PHLTA,B149,487;%%
\bibitem{Smit:1986fn}
J.~Smit and J.~C.~Vink,
Nucl. Phys. B286 (1987) 485 and
%%CITATION = NUPHA,B286,485;%%
%J.~Smit and J.~C.~Vink,
B298 (1988) 557.
%%CITATION = NUPHA,B298,557;%%
\bibitem{Boyd:1997nt}
G.~Boyd et al.,
%``Topology in QCD,''
arXiv:hep-lat/9711025.
%%CITATION = HEP-LAT 9711025;%%
\bibitem{Teper:1999wp}
M.~Teper,
%``Topology in QCD,''
Nucl. Phys. B (Proc. Suppl.)  83-84 (2000) 146. %[arXiv:hep-lat/9909124].
%%CITATION = HEP-LAT 9909124;%%
\bibitem{Lucini:2001ej}
B.~Lucini and M.~Teper,
%``SU(N) gauge theories in four dimensions: Exploring the approach to N =  infinity,''
JHEP 0106 (2001) 050. %[arXiv:hep-lat/0103027].
%%CITATION = HEP-LAT 0103027;%%
\bibitem{Cundy:2002hv}
N.~Cundy, M.~Teper and U.~Wenger,
%``Topology and chiral symmetry breaking in SU(N(c)) gauge theories,''
arXiv:hep-lat/0203030 and these proceedings.
%%CITATION = HEP-LAT 0203030;%%
\bibitem{DelDebbio:2002xa}
L.~Del Debbio, H.~Panagopoulos and E.~Vicari,
%``Theta dependence of SU(N) gauge theories,''
JHEP 0208 (2002) 044 and these proceedings. %[arXiv:hep-th/0204125].
%%CITATION = HEP-TH 0204125;%%
\bibitem{ichep2002}
G. Unal,
``Final measurement of $\varepsilon'/\varepsilon$ by NA48'', ICHEP 2002 Amsterdam;\\
T. Barker,
``Recent Results from KTeV'', ICHEP 2002 Amsterdam.
\bibitem{pdg2002}
K.~Hagiwara et al.  [Particle Data Group],
Phys. Rev. D66 (2002) 010001.
%%CITATION = PHRVA,D66,010001;%%
\bibitem{Ciuchini:2000de}
M.~Ciuchini et al.,
JHEP 0107 (2001) 013. %[arXiv:hep-ph/0012308].
%%CITATION = HEP-PH 0012308;%%
\bibitem{Hocker:2001xe}
A.~Hocker et al.,
Eur. Phys. J. C21 (2001) 225. %[arXiv:hep-ph/0104062].
%%CITATION = HEP-PH 0104062;%%
\bibitem{Ciuchini:1999xi}
M.~Ciuchini et al., arXiv:hep-ph/9910237.
%%CITATION = HEP-PH 9910237;%%
\bibitem{Bertolini:2002rj}
S.~Bertolini, 
%``Theory status of $\varepsilon'/\varepsilon$''
arXiv:hep-ph/0206095.
%%CITATION = HEP-PH 0206095;%%
\bibitem{deRafael}
E. de Rafael, these proceedings.
\bibitem{Bernard:wf}
C.~Bernard et al.,
%C.~W.~Bernard, T.~Draper, A.~Soni, H.~D.~Politzer and M.~B.~Wise,
Phys. Rev. D32 (1985) 2343.
%%CITATION = PHRVA,D32,2343;%%
\bibitem{Maiani:1986db}
L.~Maiani et al.,
%L.~Maiani, G.~Martinelli, G.~C.~Rossi and M.~Testa,
Nucl. Phys. B289 (1987) 505.
%%CITATION = NUPHA,B289,505;%%
\bibitem{Dawson:1997ic}
C.~Dawson et al.,
%``New lattice approaches to the Delta(I) = 1/2 rule,''
Nucl. Phys. B514 (1998) 313. [arXiv:hep-lat/9707009].
%%CITATION = HEP-LAT 9707009;%%
\bibitem{Maiani:ca}
L.~Maiani and M.~Testa,
%``Final State Interactions From Euclidean Correlation Functions,''
Phys. Lett. B245 (1990) 585.
%%CITATION = PHLTA,B245,585;%%
\bibitem{Lellouch:2000pv}
L.~Lellouch and M.~L\"uscher,
%``Weak transition matrix elements from finite-volume correlation  functions,''
Commun. Math. Phys. 219 (2001) 31. %[arXiv:hep-lat/0003023].
%%CITATION = HEP-LAT 0003023;%%
\bibitem{Lin:2001ek}
C.~J.~Lin et al.,
%``K $\to$ pi pi decays in a finite volume,''
Nucl. Phys. B619 (2001) 467. %[arXiv:hep-lat/0104006].
%%CITATION = HEP-LAT 0104006;%%
\bibitem{Buras:1993dy}
A.~J.~Buras, M.~Jamin and M.~E.~Lautenbacher,
Nucl. Phys. B408 (1993) 209. %[arXiv:hep-ph/9303284].
%%CITATION = HEP-PH 9303284;%%
\bibitem{Ciuchini:1993vr}
M.~Ciuchini et al., Nucl. Phys. B415 (1994) 403. %[arXiv:hep-ph/9304257].
%%CITATION = HEP-PH 9304257;%%
\bibitem{Noaki:2001un}
J.~I.~Noaki et al.  [CP-PACS Collaboration],
%``Calculation of non-leptonic kaon decay amplitudes from K $\to$ pi matrix  elements in quenched domain-wall QCD,''
arXiv:hep-lat/0108013.
%%CITATION = HEP-LAT 0108013;%%
\bibitem{Bertolini:1998vd}
S.~Bertolini, M.~Fabbrichesi and J.~O.~Eeg,
%``Theory of the CP-violating parameter epsilon'/epsilon,''
Rev. Mod. Phys. 72 (2000) 65. %[arXiv:hep-ph/9802405].
%%CITATION = HEP-PH 9802405;%%
\bibitem{Pallante:2001he}
E.~Pallante, A.~Pich and I.~Scimemi,
%``The standard model prediction for epsilon'/epsilon,''
Nucl. Phys. B617 (2001) 441. %[arXiv:hep-ph/0105011].
%%CITATION = HEP-PH 0105011;%%
\bibitem{Herrlich:1995hh}
S.~Herrlich and U.~Nierste,
%``Indirect CP violation in the neutral kaon system beyond leading logarithms,''
Phys. Rev. D52 (1995) 6505; %[arXiv:hep-ph/9507262].
%%CITATION = HEP-PH 9507262;%%
Nucl. Phys. B476 (1996) 27. %[arXiv:hep-ph/9604330].
%%CITATION = HEP-PH 9604330;%%
\bibitem{DeGrand}
T.~DeGrand, these proceedings.
\bibitem{Aoki:1997nr}
S.~Aoki et al.  [JLQCD Collaboration],
%``Kaon B parameter from quenched lattice QCD,''
Phys. Rev. Lett.  80 (1998) 5271. %[arXiv:hep-lat/9710073].
%%CITATION = HEP-LAT 9710073;%%
\bibitem{Kilcup:1997ye}
G.~Kilcup, R.~Gupta and S.~R.~Sharpe,
%``Staggered fermion matrix elements using smeared operators,''
Phys. Rev. D57 (1998) 1654. %[arXiv:hep-lat/9707006].
%%CITATION = HEP-LAT 9707006;%%
\bibitem{Aoki:1999gw}
S.~Aoki et al.  [JLQCD Collaboration],
%``The kaon B-parameter with the Wilson quark action using chiral Ward  identities,''
Phys. Rev. D60 (1999) 034511. %[arXiv:hep-lat/9901018].
%%CITATION = HEP-LAT 9901018;%%
\bibitem{Becirevic:2002mm}
D.~Becirevic et al. [SPQCDR Collaboration],
%``Kaon weak matrix elements with Wilson fermions,''
these proceedings.
\bibitem{AliKhan:2001wr}
A.~Ali Khan et al.  [CP-PACS Collaboration],
%``Kaon B parameter from quenched domain-wall QCD,''
Phys. Rev. D64 (2001) 114506. %[arXiv:hep-lat/0105020].
%%CITATION = HEP-LAT 0105020;%%
\end{thebibliography}
\end{document}